\documentclass[
reprint,           % journal's actual layout
superscriptaddress,% authors with affiliations via superscripts
amsmath,           % add AMS-Latex features
amssymb,           % add extra AMS symbols, including amsfonts
aps,               % aps or aip
prl,               % prl, pra, prb, prc, prd, pre, prstab
notitlepage,       % control appearance of title page
floatfix,          % process floats as early as possible
]{revtex4-1}

\usepackage{tensor}     % manipulate tensors
\usepackage{graphicx}   % include figures
\usepackage[
colorlinks=true,        % color link
citecolor=blue,         % cite color
linkcolor=blue,         % link color
urlcolor=blue           % url color
]{hyperref}             % create hyperlinks
\usepackage{bm}         % \bm{<text>} Bold math symbols
\usepackage{xcolor}     % textcolor
\usepackage{tabulary}
\usepackage{color}      % \textcolor{declared-color}{text}
\usepackage[section]{placeins} % figures processing
\renewcommand{\vec}[1]{\boldsymbol{#1}} % Uncomment for BOLD vectors.

\newcommand{\nc}{\newcommand*} 

\nc{\al}{\alpha}
\nc{\s}{\sigma}
\nc{\dt}{\delta}
\nc{\Dt}{\Delta}
\nc{\Ld}{\Lambda}
\nc{\p}{\partial}
\nc{\om}{\omega}
\nc{\Om}{\Omega}
\nc{\rd}{\mathrm{d}}
\nc{\Od}[1]{\mathcal{O}(#1)} % order operator
\nc{\kp}{\kappa}

\def\({\left(}
\def\){\right)}
\def\[{\left[}
\def\]{\right]}
\def\e{\begin{equation}}
\def\q{\end{equation}}
\def\m{\begin{eqnarray}}
\def\n{\end{eqnarray}}

\nc{\Eq}[1]{Eq.~\eqref{#1}}     % equation
\nc{\Fig}[1]{Fig.~\ref{#1}}     % figure
\nc{\Table}[1]{Table~\ref{#1}}  % table
\nc{\Sec}[1]{Sec.~\ref{#1}}     % section

\nc{\Msun}{M_\odot}             % solar mass
\nc{\fpbh}{f_{\mathrm{pbh}}}    % f_pbh
\nc{\fpbhn}{f_{\mathrm{pbh0}}}    % f_pbh
\nc{\mR}{\mathcal{R}} % merger rate density
\nc{\seq}{\sigma_{\mathrm{eq}}}
\nc{\ogw}{\Omega_{\mathrm{GW}}}
\nc{\gpcyr}{\mathrm{Gpc}^{-3}\,\mathrm{yr}^{-1}}
\nc{\lvc}{LIGO/Virgo} % LIGO-VIRGO collaboration
\nc{\SNR}{\mathrm{SNR}} % signal to noise ratio
\nc{\mmin}{{m_{\mathrm{min}}}}
\nc{\mmax}{{m_{\mathrm{max}}}}
\nc{\Mmin}{{M_{\mathrm{min}}}}
\nc{\fmin}{{f_{\mathrm{min}}}}
\nc{\VT}{\mathrm{VT}}
\nc{\rhoGW}{\rho_{\mathrm{GW}}}
\nc{\vth}{\vec{\theta}}
\nc{\vd}{\vec{d}}
\nc{\vla}{\vec{\lambda}}
\nc{\Nobs}{N_{\mathrm{obs}}}
\nc{\av}[1]{\langle #1 \rangle} % average bracket
\nc{\km}{\mathrm{km}}
\nc{\Mpc}{\mathrm{Mpc}}
\nc{\Tobs}{T_{\mathrm{obs}}}
\nc{\Ntemp}{N_{\mathrm{temp}}}

\nc{\addref}{[\textcolor{red}{add ref}] } % placeholder of references
\nc{\eg}{\textit{e.g.~}}
\nc{\app}{\approx}
\nc{\hf}{\frac{1}{2}}
\nc{\discuss}{\textcolor{red}{Add discussion here!}}
\nc{\red}[1]{\textcolor{red}{#1}}

\nc{\mH}{\mathcal{H}}
\nc{\cs}{c_s^2}
\nc{\Sij}[1]{S_{ij}^{(#1)}}
\nc{\vi}[1]{v_i^{(#1)}}
\nc{\no}{\nonumber}
\def\<{\left\langle}
\def\>{\right\rangle}
\def\ap{\alpha}
\def\half{{1\over 2}}
\nc{\bk}{\bm{k}}
\nc{\bq}{\bm{q}}
\nc{\bp}{\bm{p}}
\nc{\bl}{\bm{l}}
\nc{\bx}{\bm{x}}
\nc{\be}{\mathbf{e}}
\nc{\mS}{\mathcal{S}}
\nc{\te}{\tilde{\eta}}
\nc{\tp}{\tilde{p}}
\nc{\tk}{\tilde{k}}
\nc{\tx}{\tilde{x}}
\nc{\tF}{\tilde{F}}
\nc{\tA}{\tilde{A}}
\nc{\mkpq}{|\bk-\bp-\bq|}
\nc{\mpq}{|\bp-\bq|}
\nc{\mkp}{|\bk-\bp|}
\nc{\mSi}[1]{\mS^{(#1)}({\bk, \eta})}
\nc{\vk}{\vec{k}}
\nc{\kstar}{k_*}
\nc{\xstar}{x_*}
\nc{\mpbh}{m_{\rm{pbh}}}
\begin{document}
%%%%%%%%%%%%%%%%%%%%%%%%%%%%%%%%%%%%%%%%%%%%%%%%%%%%%%%%%%%%%%%%%%%%%%%%%%%%%%%%
	
%%%%%%%%%%%%%%%%%%%%%%%%%%%%%%%%%%%% title %%%%%%%%%%%%%%%%%%%%%%%%%%%%%%%%%%%%%	
\title{Probing Primordial-Black-Hole Dark Matter with Scalar Induced Gravitational Waves}
	
%%%%%%%%%%%%%%%%%%%%%%%%%%%%%%%%%%%% author %%%%%%%%%%%%%%%%%%%%%%%%%%%%%%%%%%%%
\author{Chen Yuan}
\email{yuanchen@itp.ac.cn}
\affiliation{CAS Key Laboratory of Theoretical Physics, 
	Institute of Theoretical Physics, Chinese Academy of Sciences,
	Beijing 100190, China}
\affiliation{School of Physical Sciences, 
	University of Chinese Academy of Sciences, 
	No. 19A Yuquan Road, Beijing 100049, China}
	
%%%%%%%%%%%%%%%%%%%%%%%%%%%%%%%%%%%% author %%%%%%%%%%%%%%%%%%%%%%%%%%%%%%%%%%%%
\author{Zu-Cheng Chen}
\email{chenzucheng@itp.ac.cn} %bingining@gmail.com
\affiliation{CAS Key Laboratory of Theoretical Physics, 
	Institute of Theoretical Physics, Chinese Academy of Sciences,
	Beijing 100190, China}
\affiliation{School of Physical Sciences, 
	University of Chinese Academy of Sciences, 
	No. 19A Yuquan Road, Beijing 100049, China}
	
%%%%%%%%%%%%%%%%%%%%%%%%%%%%%%%%%%%% author %%%%%%%%%%%%%%%%%%%%%%%%%%%%%%%%%%%%
\author{Qing-Guo Huang}
\email{huangqg@itp.ac.cn}
\affiliation{CAS Key Laboratory of Theoretical Physics, 
	Institute of Theoretical Physics, Chinese Academy of Sciences,
	Beijing 100190, China}
\affiliation{School of Physical Sciences, 
	University of Chinese Academy of Sciences, 
	No. 19A Yuquan Road, Beijing 100049, China}
\affiliation{Center for Gravitation and Cosmology, 
	College of Physical Science and Technology, 
	Yangzhou University, Yangzhou 225009, China}
\affiliation{Synergetic Innovation Center for Quantum Effects and Applications, 
	Hunan Normal University, Changsha 410081, China}

\date{\today}
%%%%%%%%%%%%%%%%%%%%%%%%%%%%%%%%%%%%%%%%%%%%%%%%%%%%%%%%%%%%%%%%%%%%%%
\begin{abstract}
The possibility that primordial black holes (PBHs) represent all of the dark matter (DM) in the Universe and explain the coalescences of binary black holes detected by LIGO/Virgo has attracted a lot of attention. PBHs are generated by the enhancement of scalar perturbations which inevitably produce the induced gravitational waves (GWs). We calculate the induced GWs up to the third-order correction which not only enhances the amplitude of induced GWs, but also extends the cutoff frequency from $2k_*$ to $3k_*$. Such effects of the third-order correction lead to an around $10\%$ increase of the signal-to-noise ratio (SNR) for both LISA and pulsar timing array (PTA) observations, and significantly widen the mass range of PBHs in the stellar mass window accompanying detectable induced GWs for PTA observations including IPTA, FAST and SKA. On the other hand, the null detections of the induced GWs by LISA and PTA experiments will exclude the possibility that all of the DM is comprised of PBHs and the GW events detected by LIGO/Virgo are generated by PBHs. 
		
%The possibility that the primordial black holes (PBHs) may represent the dark matter (DM) in our universe has attracted interest recently. PBHs may form from the enhancement of the scalar perturbations. These scalar perturbations will inevitably generate gravitational waves (GWs), which can be used to detect the existence of PBHs. We revisit the scalar induced GWs and extend the GWs to the fourth-order. We show that the higher-order correction of the induced GWs would not only lead to a 15\% increase of the signal to noise ratio (SNR) for both LISA and pulsar timing observation but also widen the detectable mass range of PBHs. These planned GW projects (we consider IPTA/FAST/SKA/LISA) could find or rule out PBHs in the mass ranges from $10^{-18}M_\odot$ to $10M_\odot$.
\end{abstract}
	
\pacs{???}
	
\maketitle
	
%%%%%%%%%%%%%%%%%%%%%%%%%%%%%%%%%%%%%%%%%%%%%%%%%%%%%%%%%%%%%%%%%%%%%%
%%%%%%%%%%%%%%%%%%%%%%%%%%%%%%%%%%%%%%%%%%%%%%%%%%%%%%%%%%%%%%%%%%%%%%
Various independent cosmological observations indicate the existence of dark matter (DM) in our Universe. The nature of DM remains highly elusive despite decades of dedicated searches. However, cosmological observations are only sensitive to the macroscopic properties of DM.
Even though a new elementary particle is postulated in standard DM scenarios, primordial black hole (PBH) DM has attracted a lot of attention \cite{Bird:2016dcv,Garcia-Bellido:2017fdg,Sasaki:2018dmp,Barack:2018yly},
ever since the first direct detection of gravitational waves (GWs) from a binary black hole (BBH)
coalescence \cite{Abbott:2016blz}. 
The primordial-origin BBHs are appealing candidates of \lvc\ BBHs if the abundance of stellar mass PBHs in DM is a few part in thousand \cite{Chen:2018czv,Chen:2019irf,Chen:2018rzo}. 
There are various powerful observational constraints in literature \cite{Carr:2009jm,Barnacka:2012bm,Graham:2015apa,Niikura:2017zjd,Griest:2013esa,%
Niikura:2019kqi,Tisserand:2006zx,Brandt:2016aco,Gaggero:2016dpq,%
Ali-Haimoud:2016mbv,Blum:2016cjs,Horowitz:2016lib,Chen:2016pud,%
Wang:2016ana,Abbott:2018oah,Magee:2018opb,Wang:2019kaf,Chen:2019irf,Montero-Camacho:2019jte,Laha:2019ssq}, but a substantial window remains open for PBHs as all of DM in the approximate range $[10^{-16},10^{-14}] \cup [10^{-13},10^{-12}] M_\odot$. See a recent summary in \cite{Chen:2019irf}.

PBHs are supposed to form from the enhancement of the scalar perturbations \cite{Hawking:1971ei,Carr:1974nx}. The process during which the PBHs are formed would be inevitably accompanied by GWs \cite{Tomita1967,Matarrese:1992rp,Matarrese:1993zf,Matarrese:1997ay,Noh:2004bc,Carbone:2004iv,Nakamura:2004rm,Ananda:2006af}.
These so-called induced GWs are driven by scalar perturbations during radiation-dominated (RD) era and could leave detectable signal at present for testing the hypothesis of PBH DM  \cite{Saito:2008jc,Bugaev:2010bb,Sasaki:2018dmp,Inomata:2018epa,Baumann:2007zm,Clesse:2018ogk,Nakama:2016enz,Saito:2009jt,Bugaev:2009zh,Assadullahi:2009jc}. 
Recently, a semi-analytical expression for the GWs induced by second-order scalar perturbations
has been derived in \cite{Espinosa:2018eve,Kohri:2018awv}, and the discussions have also been extended to detect the primordial non-Gaussianity 
\cite{Garcia-Bellido:2017aan,Unal:2018yaa,Cai:2018dig,Cai:2019amo} 
through induced GWs and the bispectrum
\cite{Bartolo:2018evs,Bartolo:2018rku,Espinosa:2018eve} of the induced GWs.
PBHs could have been produced during RD era when relatively large scalar perturbations with amplitudes $\mathcal{O}(0.01-0.1)$ re-entered the Hubble horizon \cite{Ivanov:1994pa,GarciaBellido:1996qt,Ivanov:1997ia,Yokoyama:1995ex,Kawasaki:2006zv}. The non-linearity is an intrinsic property for gravity in general relativity, and therefore the higher-order corrections to the scalar induced GWs are expected to be important. 
%, rendering higher-order corrections of the induced GWs may play a non-negligible role in observations. 
However, previous studies only focused on the GWs induced by second-order scalar perturbations, 
and as far as we know, no higher-order effects have been taken into account in literature.
In this letter, we give the first result of the induced GWs up to the third-order, and find that the third-order correction would be detectable by some future GW detectors, such as LISA \cite{Audley:2017drz}, IPTA \cite{Hobbs:2009yy}, FAST \cite{Nan:2011um} and SKA \cite{Kramer:2015jsa}, and has significant observational implications.
%By considering LISA and PTA observations including IPTA/FAST/SKA, we show that the third-order correction is detectable and thus, indispensable in making more solid constraints on PBHs by these future GW projects.
	
The perturbed metric in the FRW spacetime with Newtonian gauge takes the form, \cite{Ananda:2006af}, 
\e
\rd s^{2}=a^{2}\left\{-(1+2 \phi) \rd \eta^{2}+\left[(1-2 \phi) \delta_{i j}+\frac{h_{i j}}{2}\right] \rd x^{i} \rd x^{j}\right\},
\q
where $\phi$ and $h_{ij}$ are the scalar and tensor perturbations respectively.
%The vector and the anisotropic stress perturbations have been neglected \cite{Baumann:2007zm}. Here, we will be only interested in RD era and hence set the equation of state $w$ and acoustic speed $c_s$ to $w = c_s^2 = 1/3$ (results without fixing $w$ and $c_s$ will be available in \cite{Yuan:2019}).
The scalar perturbation in Fourier space has a solution \cite{Baumann:2007zm,Kohri:2018awv,Sasaki:2018dmp,tomita1971}
\e\label{phi}
	\phi_{\vec{k}}(\eta) \equiv \phi_{\vec{k}}T(k\eta),
\q
where $\phi_{\vec{k}}$ is the primordial perturbation and $T(k\eta)$ is the transfer function
\e\label{transfer}
	T(k\eta) =  \frac{9}{(k\eta)^2} \[ \frac{\sin(k\eta/\sqrt{3})}{k\eta/\sqrt{3}} - \cos(k\eta/\sqrt{3})\],
\q
which oscillates and decays as $\sim1/\eta^2$ in the RD era. 
In order to extract the induced GWs up to the third-order, it is necessary to expand Einstein equations up to the fourth-order 
(perturbations are performed utilizing the \texttt{xPand} \cite{Pitrou:2013hga} package). 
After a tedious but straightforwards computation, the evolution of the GWs up to the fourth-order is given by
\e\label{eqh}
h_{i j}^{\prime \prime}+2 \mathcal{H} h_{i j}^{\prime}-\nabla^{2} h_{i j}=-4 \mathcal{T}_{i j}^{\ell m} S_{\ell m},
\q
where the prime denotes the derivative with respect to the conformal time $\eta$, 
$\mH \equiv a'/a$ is the conformal Hubble parameter, and 
$\mathcal{T}_{i j}^{\ell m}$ is the projection operator \cite{Ando:2017veq} 
onto the transverse and traceless tensor.
Although \Eq{eqh} has the same form as the evolution of GWs at second-order 
(see \eg \cite{Ando:2017veq,Baumann:2007zm}),
the source term $S_{ij}=\Sij{2}+\Sij{3}+\Sij{4}$ has been calculated up to fourth-order as follows 
%is obtained by extending the energy-momentum tensor to fourth-order. In the derivation of the  source terms, one would expect to see that the higher-order GWs are driven by not only the same order perturbations but also the lower order GWs. However, as the amplitude of the GWs is far smaller than the scalar perturbations, it is robust to keep only the scalar perturbations. The source terms in real space up to fourth-order reads
\e\label{S2}
\Sij{2}= 4 \phi \p_i\p_j\phi + 2\p_i\phi\p_j\phi-\p_i \(\phi + {\phi'\over\mH}\)
\p_j\(\phi + {\phi'\over\mH}\),
\q
%\m
%\Sij{2} = 4 \phi \phi_{ij} + 2\phi_i\phi_j-\frac{1}{\mH} \(\phi_i + { \phi_i'\over\mH}\)
%\(\phi_j + {\phi_j'\over\mH}\),
%\n	
\e\label{S3}
\begin{split}
\Sij{3} =& \frac{1}{\mH} \(12\mH \phi - \phi'\) \p_i\phi \p_j\phi
	- \frac{1}{\mH^3} \(4\mH \phi - \phi'\)\p_i\phi'\p_j\phi' \\
	&+ \frac{1}{3\mH^4} \(2\p^2\phi - 9\mH \phi'\) \p_i\(\mH\phi +\phi'\)
	\p_j\(\mH\phi +\phi'\),
\end{split}
\q
%\m
%\Sij{3} &&= \frac{1}{\mH} \(12\mH \phi - \phi'\) \phi_i \phi_j
%- \frac{1}{\mH^3} \(4\mH \phi - \phi'\)\phi_i'\phi_j'\no\\
%&&+ \frac{1}{3\mH^4} \(2\p^2\phi - 9\mH \phi'\) \(\mH \phi_i + \phi_i'\)
%\(\mH\phi_j + \phi_j'\),
%\n
\e\label{S4}
\begin{split}
\Sij{4} =& 16 \phi^3 \p_i\p_j\phi
+ \frac{1}{3\mH^3} \Big[2 \phi' \p^2\phi - 9\mH \phi'^2 
- 8\mH \phi \p^2\phi \\
&\qquad\qquad\qquad\quad+ 18 \mH^2 \phi \phi'
+ 96\mH^3 \phi^2\Big] \p_i\phi\p_j\phi \\
&+ \frac{2}{3\mH^5} \Big[- \phi' \p^2\phi + 3\mH \phi'^2 
+ 4\mH \phi \p^2\phi \\
&\qquad\qquad\qquad\quad\quad+ 3 \mH^2 \phi \phi'- 12\mH^3 \phi^2\Big] \p_i\phi' \p_j\phi'\\
&+ \frac{1}{36\mH^6} \Big[ 
-16 (\p^2\phi)^2 - 3  \partial_k\phi' \partial^k\phi'
+ 120 \mH \phi' \p^2\phi \\
&\qquad\quad - 6 \mH \partial_k\phi \partial^k\phi'	+ 144 \mH^2 \phi \p^2\phi - 180 \mH^2 \phi'^2 \\
&\qquad\qquad + 33 \mH^2 \partial_k\phi \partial^k\phi- 504 \mH^3 \phi \phi'
- 144 \mH^4 \phi^2
\Big] \\
&\qquad\qquad\times  \p_i\(\mH\phi +\phi'\)
\p_j\(\mH\phi +\phi'\).
\end{split}
\q
%Previous studies only focused on GWs induced by $\Sij{2}$ (see \eg \cite{Ando:2017veq,Baumann:2007zm}) and ignored higher-order effects.
%The expressions for $\Sij{3}$ and $\Sij{4}$ are derived here for the first time, and the effects of these terms should not be safely neglected as we will show.

%where in Fourier space, the perturbations $\phi_{\vec{k}}(\eta)$ will decay as $\sim1/\eta^2$ (Eq.~\ref{phi}) so that the source terms will also tend to oscillate and decay during RD era.
%A genearl expression of $\Sij{3}$ and $\Sij{4}$ will be given in future work.
	
After solving Eq.~(\ref{eqh}) in Fourier space by Green's function, we can use the two-point correlation of the GWs to calculate their power spectrum, namely 
\e\label{Ph}
	\left\langle h(\vec{k},\eta) h(\vec{k'},\eta)\right\rangle\equiv\frac{2 \pi^{2}}{k^{3}} \mathcal{P}_{h}(k,\eta) \delta(\vec{k}+\vec{k'}).
\q
The angle brackets stand for ensemble average which can be calculated using timing average instead. The GW energy density, $\rho_{\mathrm{GW}}=\int\rho_{\mathrm{GW}}(k,\eta)\, \rd\ln k$, 
can be evaluated as \cite{Maggiore:1999vm} 
\e
\rho_{\mathrm{GW}}={M_p^2\over16a^2}\<\overline{\partial_kh_{ij}\partial^kh^{ij}}\>,
\q
where the overline stands for the average of the oscillating effect of the time-varying phase and $M_p$ is the Planck mass. The dimensionless GW energy density parameter $\ogw$ is defined as the energy density of GWs per logarithmic frequency normalized by the critical density $\rho_{\rm{cr}}$,
\e\label{omiga}
\ogw(\eta, k)\equiv\frac{\rho_{\mathrm{GW}}(k,\eta)}{\rho_{\rm{cr}}} = \frac{1}{12} \(\frac{k}{\mH}\)^2 \overline{\mathcal{P}_{h}(k, \eta)},
\q
where we have summed over the two polarization modes of $+$ and $\times$.
Solving \Eq{eqh} by the Green's function method, one obtains \cite{Baumann:2007zm}
\e\label{hsol} 
	h(\vec{k},\eta) = \frac{1}{k a(\eta)} \int \rd \te \sin(k\eta - k\te) a (\te) \mathcal{S}_{\vec{k}}(\te),
\q 
where $\mathcal{S}_{\vec{k}}(\eta)\equiv -4e^{ij}(\vec{k}) \tilde{S}_{ij}(\vec{k},\eta)$ with $\tilde{S}_{ij}(\vec{k},\eta)$ being the source term transformed into Fourier space.
The polarization tensors $e_{ij}(\vec{k})$ are defined as 
$(e_i e_j - \bar{e}_i \bar{e}_j)/\sqrt{2}$ and $(e_i \bar{e}_j + \bar{e}_i e_j)/\sqrt{2}$ 
for $+$ and $\times$ polarizations respectively, 
where $e_i(\vk)$ and $\bar{e}_i(\vk)$ are two independent unit vectors orthogonal to $\vk$.
Since $\phi$ is related to the comoving curvature perturbation $\zeta$ by $\phi = (2/3) \zeta$ on superhorizon scales, $\ogw(\eta, k)$ can be calculated using the power spectrum of the comoving curvature perturbation, $\mathcal{P}_{\zeta}(k)$, defined by 
\e
\left\langle \zeta_{\vk} \zeta_{\vec{k'}}\right\rangle\equiv\frac{2 \pi^{2}}{k^{3}} \mathcal{P}_{\zeta}(k) \delta(\vec{k}+\vec{k'}). 
\q

From now on we will dedicated to the following monochromatic power spectrum
\e\label{pzeta}
	\mathcal{P}_{\zeta}(k)=A \kstar\delta\(k-\kstar\),
\q
to illustrate the effects of third-order correction.
Here $A$ is an overall normalization coefficient and $\kstar$ is the wavenumber at which the
power spectrum has a $\delta$-function peak.
Note that it is robust to neglect the contribution of the long wavelength modes on the CMB scales,
because not only the time delay effect of the long wavelength will not affect the GWs power spectrum \cite{Bartolo:2018rku}, but also the long wavelength mode is well outside the horizon during the formation of PBHs and thus should not affect any local physical processes. 
On the other hand, the $\delta$-spectrum also corresponds to a monochromatic PBH formation. 
The formation of PBHs is a threshold process where the comoving curvature perturbation $\zeta{\vec{(x)}}$ exceeds a threshold value $\zeta_c$ and causes an overdensed region. The possibility of forming a PBH can be evaluated statistically by integrating the probability density function over the threshold region \cite{Carr:2016drx}
\e\label{beta}
	\beta=\int_{\zeta_{c}}^{+\infty} \frac{\mathrm{d} \zeta}{\sqrt{2 \pi} \sigma} e^{-\zeta^{2} / 2 \sigma^{2}}=\half\mathrm{erfc}\({\zeta_{c}\over\sqrt{2A}}\),
\q
where $\zeta_{c}\simeq1$ is the threshold value \cite{Musco:2008hv,Musco:2004ak,Musco:2012au,Harada:2013epa} to form a PBH and $\sigma^2\equiv\<\zeta^2\>=\int \mathcal{P}_\zeta(k) \mathrm{d}\ln k =A$ is the variance of the curvature perturbation. % which is related to the dimensionless amplitude of its power spectrum.
For monochromatic PBHs, the possibility to form a single PBH, $\beta$, is equivalent to the abundance of PBHs which is related to the fraction of PBHs by \cite{Nakama:2016gzw}
\m\label{fpbh}
	f_{\mathrm{pbh}}\simeq 2.5 \times 10^{8}\beta\left(\frac{g_*^{\mathrm{form}}}{10.75}\right)^{-\frac{1}{4}}\left(\frac{m_{\mathrm{pbh}}}{M_{\odot}}\right)^{-\frac{1}{2}},
\n
where $g_*^{\mathrm{form}}$ is the effective degrees of freedom when PBHs are formed, 
%Once a overdensed region is formed during inflation, it would collapse to form a PBH immediately after the corresponding wavelength, $k$, re-enters the horizon so that the mass of the PBH roughly equals to the horizon mass which reads \cite{Hawking:1971ei,Carr:1974nx}
%(using $k=aH=2\pi f$), 
and the mass of the PBH roughly equals to the horizon mass, namely 
\m\label{mkrelation}
	{m_{\mathrm{pbh}}\over M_{\odot}}\approx2.3\times10^{18}\left(\frac{3.91}{g_*^{\mathrm{form}}}\right)^{1 / 6}\left(\frac{H_{0}}{f_*}\right)^{2},
\n
where we have used $k_*=aH_*=2\pi f_*$ and $H_0$ is the Hubble constant at present.

For the $\delta$-spectrum in \Eq{pzeta}, the source terms in Fourier space are given by
\m
	\mathcal{S}^{(2)}_{\vec{k}}(\eta)&\equiv&\int \frac{\rd^3p}{(2\pi)^{3/2}} \be(\bp, \bp)
	f_2(\kstar \eta) \zeta_{\bp} \zeta_{\bm{k-p}},\\
	\mathcal{S}^{(3)}_{\vec{k}}(\eta)&\equiv&\int \frac{\rd^3p \rd^3q}{(2\pi)^3} \be(\bp, \bq)
	f_3(\kstar \eta) \zeta_{\bp} \zeta_{\bq} \zeta_{\bm{k-p-q}},\\
	\mathcal{S}^{(4)}_{\vec{k}}(\eta)&\equiv&\int \frac{\rd^3p \rd^3q \rd^3l}{(2\pi)^{9/2}}\[ 
	\be(\bl, \bl)+ \be(\bp,\bq)\] f_4(\kstar \eta) \no\\
	&&\qquad\qquad\qquad \times \zeta_{\bp} \zeta_{\bq} \zeta_{\bl} \zeta_{\bm{k-p-q-l}},
\n
where we have defined $\be(\bp, \bq) \equiv e^{ij}(\vec{k})p_iq_j$, and $f_i(x)$ ($i = 2, 3, 4$) 
have the following functional forms
\m
	f_2(x)&&={8\over9}\(3T^2 + 2xTT' + x^2T'^2\),\\
	f_3(x)&&= -\frac{64}{81} \Big[ \(x^2-18\) T^3 + 2x \(3+x^2\) T^2 T'
	\no\\
	&&\qquad\quad + x^2 \(15 + x^2\) T T'^2 + 3 x^3 T'^3 \Big], \\
	f_4(x)&&= {16\over729}\Big[\(720-29x^2+16x^4\)T^4\no\\
	&&\qquad\quad+4x\(144+73x^2+8x^4\)T^3T'\no\\
	&&\qquad\quad+2x^2\(864+219x^2+8x^4\)T^2T'^2\no\\
	+4&&x^3\(198 + 31 x^2\)TT'^3+x^4 \(108 + 7 x^2\) T'^4\Big].
\n
The explicit expression for transfer function $T=T(x)$ can be found in \Eq{transfer}. 
From Eqs.~\eqref{Ph} and \eqref{hsol}, we see that only $\av{\mathcal{S}^{(2)}_{\vec{k}} \mathcal{S}^{(2)}_{\vec{k'}}}$ contributes to second-order induced GWs.
Meanwhile, both $\av{\mathcal{S}^{(3)}_{\vec{k}} \mathcal{S}^{(3)}_{\vec{k'}}}$ 
and $\av{\mathcal{S}^{(2)}_{\vec{k}} \mathcal{S}^{(4)}_{\vec{k'}}}$ contribute to the third-order correction. 
Following the pioneering work of \cite{Espinosa:2018eve,Kohri:2018awv}, for the $\dt$-spectrum in \Eq{pzeta}, we obtain 
\e\label{ogw}
	\ogw(\eta, k)=\frac{A^2}{192\tk^{2}}\Big[\overline{I_2^2}M_1+
	A \(M_2\overline{I_3^2}+M_1\overline{I_2 I_4}\)\Big],
\q
where an overbar denotes the oscillation average \cite{Kohri:2018awv} and  $I_i$ ($i = 2, 3, 4$) are defined as
\e\label{Ii}
	I_i = \lim_{x \to \infty} \int_0^x \rd\tx\, f_i\!\!\(\frac{\tx}{\tk}\) 
	{\tx\over x} \sin(x-\tx),
\q
which reflects the phase oscillation of the GWs. 
For convenience, we have defined some dimensionless parameters, i.e. $\tk\equiv k/\kstar$, $x\equiv k\eta$ and $\tx\equiv k\tilde{\eta}$. 
Similar to \cite{Kohri:2018awv}, \Eq{Ii} can be analytically integrated by multiple usages of
the trigonometric addition theorem and the properties of sine integral $\rm{Si}(\theta)$ 
and cosine integral $\rm{Ci}(\theta)$.
%The final analytical results of $I_i$ can be found in \cite{Yuan:2019}.
The angle integrals $M_1$ and $M_2$ in \Eq{ogw} are defined as
\m
	M_1(k) &=& \(4 - \tk^2\)^2 \Theta(2 - \tk),\\
	M_2(k) &=& \frac{1}{\pi^2} \int_{p_{\rm{min}}}^{p_{\rm{max}}} \rd\tp
\int_{0}^{2\pi} \rd\ap \int_{0}^{2\pi} \rd\phi M_0 \Theta(\Delta),
\n
where $\Theta$ is the Heaviside step function,
$p_{\rm{min}} = |1-\tk|$, and $p_{\rm{max}} = \min(2,1+\tk)$.
Due to its complexity, the expression of $M_2$ is evaluated numerically and 
the definition of $M_0$ and $\Delta$ are given by 
\m
\Delta =&& 4\mu^2+4(1-\mu^2)\cos(\al-\phi)^2-\tilde{p}^2,\\
M_0 =&& \sum_{i=1}^{2} \frac{(1-\nu_i^2)}{\left|\mu \sqrt{1-\nu_i^2} 
	- \nu_i \sqrt{1-\mu^2} \cos(\ap-\phi)\right|}\no\\
\Big[ 
(1-\nu_i^2&&)^{3\over2} \cos^22\al + 2\tp^3 (1-\mu^2)^{3\over2} \cos2\phi \cos(\al+\phi)\no\\
&&- 2 \tp (1-\nu_i^2) (1-\mu^2)^{\half} \cos2\al\cos(\al+\phi)\no\\
&& - \tp^2 (1-\mu^2) (1-\nu_i^2)^{\half} \cos(\al+\phi)^2 \Big],
\n
where $\mu$ and $\nu_i$ ($i=1,2$) are defined as
\m
\mu &&= {\tp^2+\tk^2-1 \over 2\tp \tk}, \\
\nu_{1,2} &&= \frac{\tp\mu\pm\sqrt{1-\mu^2}|\cos(\al-\phi)|\sqrt{\Delta}}{2\((1-\mu^2)\cos(\al-\phi)^2+\mu^2\)}.
\n
Note that $M_1$ is originated from the 
$\av{\mathcal{S}^{(2)}_{\vec{k}} \mathcal{S}^{(2)}_{\vec{k'}}}$ and 
$\av{\mathcal{S}^{(2)}_{\vec{k}} \mathcal{S}^{(4)}_{\vec{k'}}}$ terms, 
and indicates a cutoff frequency at $k=2\kstar$;
while $M_2$ is originated from the 
$\av{\mathcal{S}^{(3)}_{\vec{k}} \mathcal{S}^{(3)}_{\vec{k'}}}$,
and indicates a cutoff at $k=3\kstar$.
Therefore the third-order correction not only enhances the amplitude of GW energy density,
but also extends the cutoff frequency from $2\kstar$ to $3\kstar$.

\begin{figure}[htbp!]
	\centering
	\includegraphics[width = 0.5\textwidth]{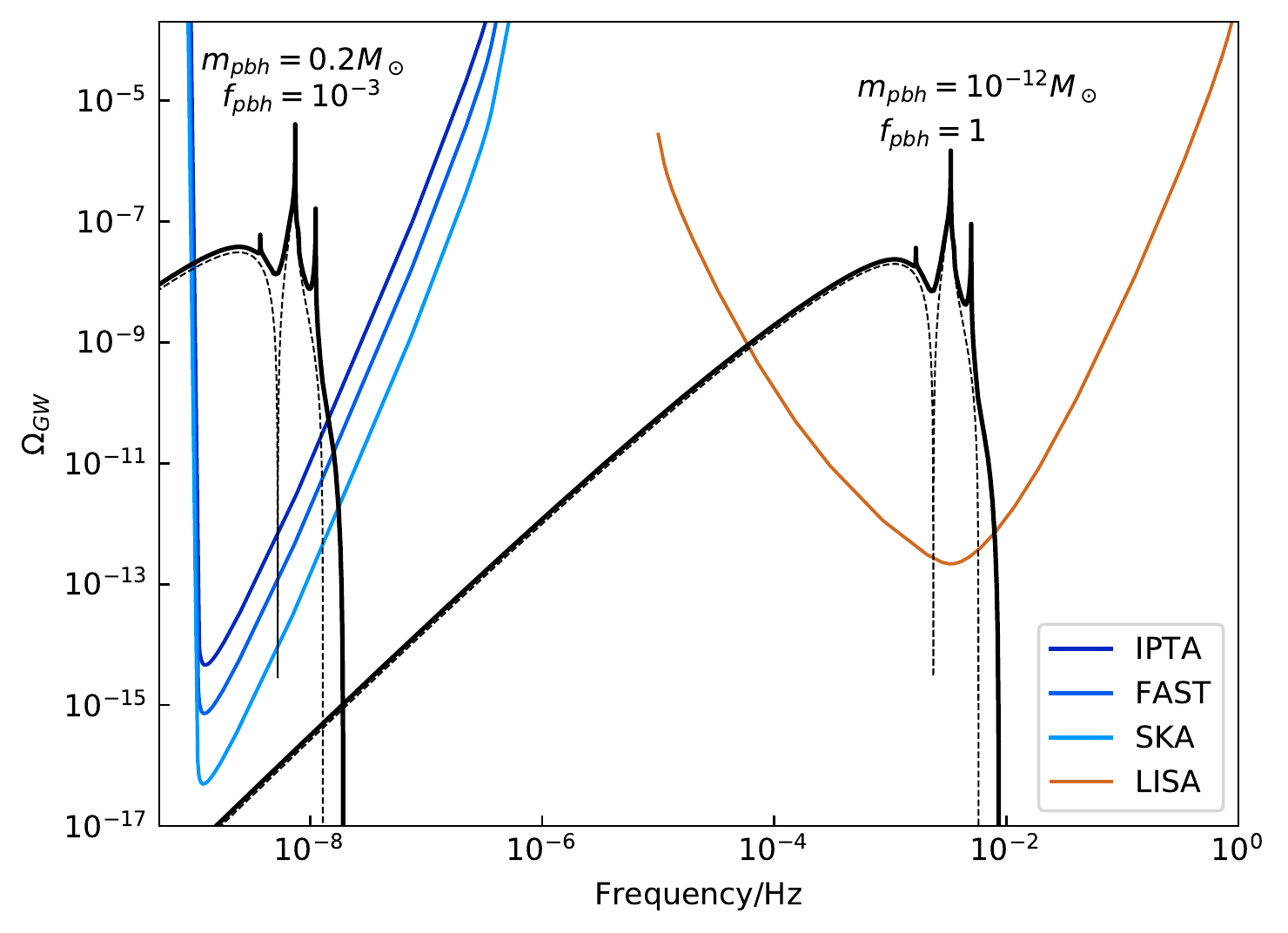}
	\caption{\label{OmegaGW}The GW density parameter of scalar induced GWs along with the power-law integrated sensitivity curves \cite{Thrane:2013oya} of LISA, IPTA, FAST and SKA. 
		%	For the pulsar timing observation, we assume pulsar timing measurements every two weeks and a 20 yr observation. 
		The black solid (dashed) lines represent $\ogw$ induced by scalar perturbations up to third-order (second-order).
		We assume IPTA, FAST and SKA last for the same observation time of $30$ years,
		and other settings of these PTA projects can be found in Table 5 of \cite{Kuroda:2015owv}.
%		The third-order correction will not only give rise to two resonant peaks at $f=(1/\sqrt{3})f_*$ and $f=\sqrt{3}f_*$ but also smooth the valley at $f=(\sqrt{2/3})f_*$.
	}
\end{figure}

Note that \Eq{ogw} is only valid from the horizon re-entry to matter-radiation equality.
Because the energy density of GWs decays as radiation, current density parameter of GWs 
can be approximated by \cite{Espinosa:2017sgp}
\e
	\ogw(\eta_0, f) \simeq \Omega_{r} \times \ogw(\eta, f),
\q
where $\Omega_{r}$ is the density parameter of radiation at present.
%It has already been noticed that the scalar-induced GWs may be strong enough to be observed by various GW experiments, or otherwise they may be used to constrain the model parameters, such as the abundance of PBHs in DM $\fpbh$.
From \Eq{mkrelation}, heavier masses of PBHs corresponds to lower peak frequencies of induced GWs. 
%Because PBHs lighter than $\sim10^{-18} \Msun$, which corresponds to frequency of $\sim3$ Hz, have already evaporated by Hawking radiation, the scalar-induced GWs will not be detect by \lvc, which are sensitive to frequencies larger than $\sim 10$ Hz. Nevertheless, these GWs may be detected by LISA or PTA observations.
\Fig{OmegaGW} shows the GWs induced by scalar perturbation up to third order compared with
the sensitivity curves of LISA, IPTA, FAST and SKA. Because a substantial windows for PBH as all of DM in the approximate range $[10^{-16},10^{-14}]\cup [10^{-13},10^{-12}] M_\odot$ is still available, for example, we choose $\mpbh=10^{-12}\Msun$ and $\fpbh=1$ and then the peak frequency of the induced GW is $f\sim 10^{-3}$ Hz which is within the LISA frequency band. In addition, roughly speaking, since the abundance of stellar mass PBHs has been constrained to be less than $0.01$ \cite{Chen:2019irf}, we choose $\mpbh=0.2\Msun$ and $\fpbh=10^{-3}$, and then the peak frequency of induced GWs is just located at the PTA frequency band. 
In fact, since Eqs.~\eqref{beta} and \eqref{fpbh} imply $f_{\rm{pbh}} \propto\sqrt{A} \exp(-A^{-1})$ for a relatively small value of $A$, 
even decreasing $\fpbh$ by multiple orders, the consequent value of $A$ will almost not change.
Therefore the amplitude of $\ogw$ will almost be irrespective with the choice of $\fpbh$ unless $\fpbh$ is dramatically much smaller than our choice in \Fig{OmegaGW}. 
From \Fig{OmegaGW}, we also see that the third-order correction not only enhances the magnitude of
$\ogw$ (especially smooth the deep valley at $f=\sqrt{2/3}f_*$ from the second-order effect), but also extends cutoff frequency from $2f_*$ to $3f_*$. 
Besides, for a narrow spectrum, it would also generate two new resonant peaks at $f=(1/\sqrt{3})f_*$ and $f=\sqrt{3}f_*$. 
%Moreover, these features make the signal distinguishable from other stochastic GW background such as binary black hole mergers.
Moreover, the typical $\ogw(\eta_0, f)$ in \Fig{OmegaGW} shows that the third-order correction of the induced GWs is also far beyond the sensitivity curves for both LISA and PTA observations, rendering third-order correction being detectable and thus, indispensable in testing the hypothesis of PBHs.

\begin{figure}[htbp!]
	\centering
	\includegraphics[width = 0.5\textwidth]{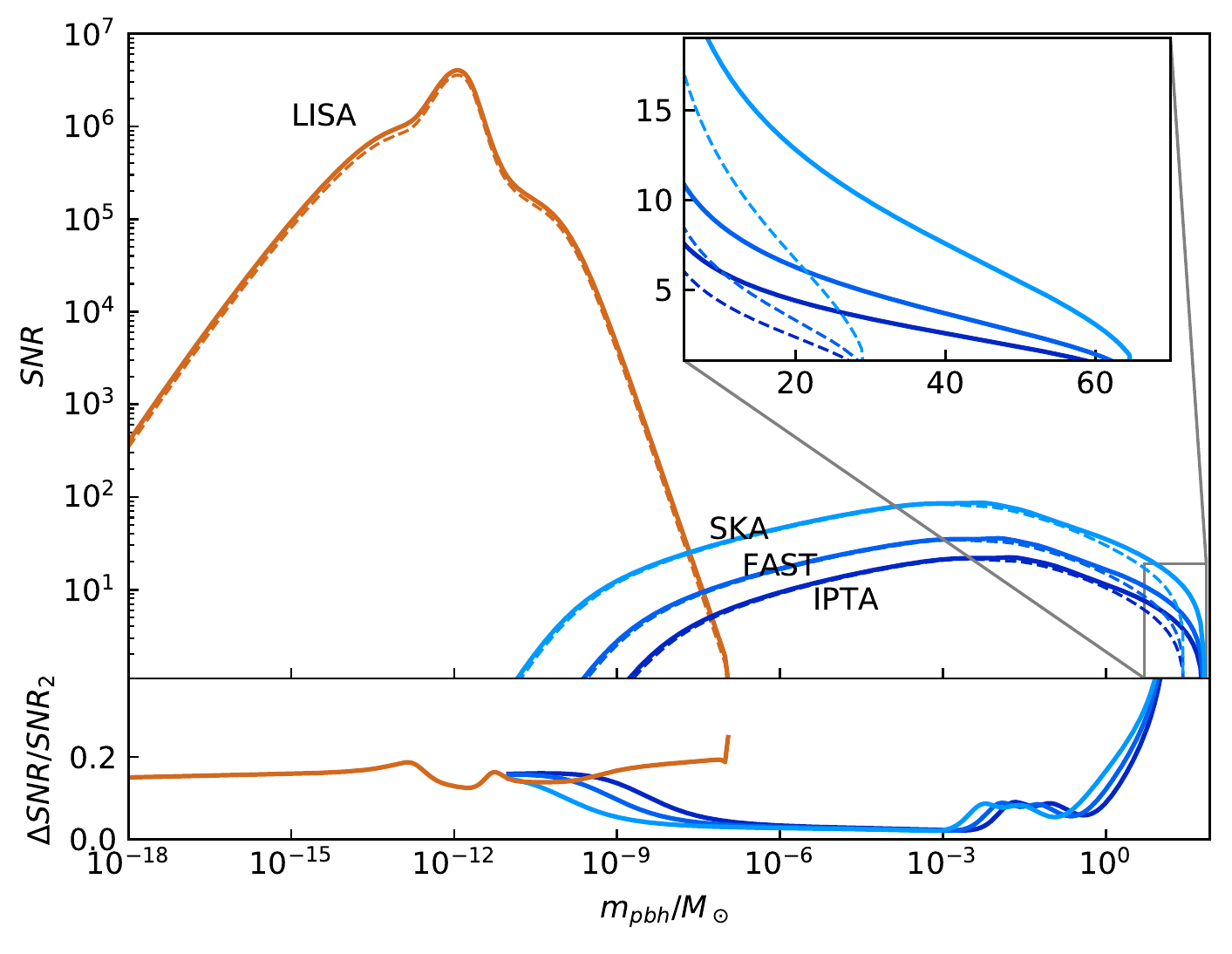}
	\caption{\label{SNR}The upper panel shows the expected SNR of the scalar-induced GWs as a function of $m_{\rm{pbh}}$ for LISA, IPTA, FAST and SKA. The solid and dashed lines represent the SNR including the third-order correction and only the second-order effects, respectively. The lower panel shows the relative change of the SNR after taking into account the third-order correction.
	}
\end{figure}

In order to quantitatively evaluate the effects of the third-order correction in observations, 
we need to estimate the SNR, $\rho$, for different GW experiments.
For LISA, it is given by \cite{Thrane:2013oya}
\e 
	\rho = \sqrt{T} \[ \int \rd f \(\frac{\ogw(f)}{\Om_n(f)}\)^2 \]^{1/2},
\q
where $\Om_n(f) = 2\pi^2 f^3 S_n/(3H_0^2)$
and $S_n$ is the strain noise power spectral density \cite{Cornish:2018dyw}. 
For PTA observations, if we assume pulsars are distributed homogeneously
and all pulsars have the same noise characteristics, the SNR is given by \cite{Siemens:2013zla} 
\e 
	\rho = \sqrt{2T} \(\sum_{I, J}^{M} \chi_{IJ}^2\[\int \rd f \(\frac{\ogw(f)}{\Om_n(f) + \ogw(f)}\)^2\]\)^{1/2},
\q
where $\chi_{IJ}$ is the Hellings and Downs coefficient for pulsars $I$ and $J$ \cite{Hellings:1983fr}.
\Fig{SNR} shows the expected SNR obtained for LISA and PTA experiments,
and indicates that the third-order correction would raise the SNR as expected. 
For LISA, we expect around a $15\%$ increase in the relative SNR;
while for PTA observations, we expect the increase is more than $5\%$ and could even reach $20\%$ for stellar mass PBHs. 
%The contribution of the third-order correction becomes important at the maximum and minimum detectable mass where the SNR is low. 
Most importantly, since the third-order correction will extend the cutoff frequency from $2k_*$ to $3k_*$, the induced GWs accompanying heavier PBHs are supposed to be detected by PTA observations. 
%For instance, in the mass range of the stellar-mass black hole, only considering the second-order effects correspond to $\sim30M_\odot$ and would leave out the possibility of detecting heavier PBHs $\sim65M_\odot$. 
For instance, as shown in \Fig{SNR}, the third-order correction will extend the largest detectable mass of PBHs from around $30\Msun$ to $65\Msun$, which will be an invaluable complimentary tool to
test the PBH scenario in addition to the analysis \lvc.

%Moreover, increasing the observation time of the observations would extend the maximum detectable mass. Under this circumstance, the third-order correction would be expected to examine heavier PBHs in the interested stellar mass window.
	
In this letter, we compute the third-order correction to the induced GWs generated by the scalar perturbations accompanying the formation of PBHs during the RD era. 
After deriving a general expression for the GW energy density, we investigate an infinite narrow power spectrum of the scalar perturbation and obtain a semi-analytical expression for $\ogw$. 
Our result implies that the third-order correction to the induced GWs accompanying formation of PBH DM will generate detectable effects on the waveform and the amplitude in observation data. 
The third-order correction not only enhances the magnitude of $\ogw$, but also extends the cutoff frequency from $2f_*$ to $3f_*$. 
We also forecast the SNR for LISA and PTA observations, including IPTA, FAST and SKA. These planned GW projects cover a wide frequency band from $10^{-9}$ Hz to $10^{-3}$ Hz corresponding to PBHs with mass range $10^{-18}M_\odot-10^{2}M_\odot$. Our results indicate that the third-order correction could not only lead to an increase in the relative SNR around $10\%$ for these planned GW projects, but also extend the maximum detectable PBH mass. On the other hand, if all of these projects are unable to detect such induced GWs, we could rule out PBH DM in a wide mass ranges of $\sim10^{-18}M_\odot-10M_\odot$. 
	
%%%%%%%%%%%%%%%%%%%%%%%%%%%%%%%%%%%%%%%%%%%%%%%%%%%%%%%%%%%%%%%%%%%%%%%%%%%%%%%%
%%%%%%%%%%%%%%%%%%%%%%%%%%%%%%% acknowledgments %%%%%%%%%%%%%%%%%%%%%%%%%%%%%%%
Acknowledgments. 
	We would like to thank Lu Chen, Yun Fang, Fan Huang, Jun Li, Lang Liu, Shi Pi, 
	You Wu, Yu Sang, Sai Wang, Hao Wei and Xue Zhang 
	for useful conversations. CY is indebted to Yu-Jia Zhai for her continuous support and encouragement.
	%We also thank the anonymous referee for valuable suggestions and comments.
	We acknowledge the use of HPC Cluster of ITP-CAS. 
	This work is supported by grants from NSFC 
	(grant No. 11690021, 11575271, 11747601), 
	the Strategic Priority Research Program of Chinese Academy of Sciences 
	(Grant No. XDB23000000, XDA15020701), and Top-Notch Young Talents Program of China.
	This research has made use of \texttt{GWSC.jl} \cite{gwsc} package to calculate
	the SNR for various gravitational-wave detectors.

%%%%%%%%%%%%%%%%%%%%%%%%%%%%%%%%%%%%%%%%%%%%%%%%%%%%%%%%%%%%%%%%%%%%%%%%%%%%%%%%
%%%%%%%%%%%%%%%%%%%%%%%%%%%%%%% %%% references %%%%%%%%%%%%%%%%%%%%%%%%%%%%%%%%%%

\end{document}